\documentclass[a4paper,12pt]{iopart}

\usepackage{graphicx}
\usepackage{dcolumn}
\usepackage{bm}
\usepackage{hyperref}
\usepackage{color}
\usepackage{ulem}
\usepackage{iopams}
\usepackage{setstack} 

\newcommand{\eqref}[1]{(\ref{#1})}

\begin{document}

\title[
Impurity-induced vector spin chirality and anomalous Hall effect
]{
Impurity-induced vector spin chirality and anomalous Hall effect in ferromagnetic metals
}

\author{Hiroaki Ishizuka$^1$ and Naoto Nagaosa$^{1,2}$}
\address{
$^1$Department of Applied Physics, The University of Tokyo, Bunkyo, Tokyo, 113-8656, JAPAN 
}
\address{
$^2$RIKEN Center for Emergent Matter Sciences (CEMS), Wako, Saitama, 351-0198, JAPAN
}
\ead{ishizuka@appi.t.u-tokyo.ac.jp}

\begin{abstract}
Scattering by multiple scatterers sometimes gives rise to nontrivial consequences such as anomalous Hall effect. We here study a mechanism for anomalous Hall effect which originates from the correlation of nonmagnetic impurities and localized moments; a Hall effect induced by vector spin chirality. Using a scattering theory approach, we study the skew scattering induced by the scattering processes that involve two magnetic moments and a non-magnetic impurity, which is proportional to the vector spin chirality of the spins in the vicinity of the non-magnetic impurity. Furthermore, we show that a finite vector spin chirality naturally exists around an impurity in the usual ferromagnetic metals at finite temperature due to the local inversion-symmetry breaking by the impurity. The result is potentially relevant to magnetic oxides which the anomalous Hall effect is enhanced at finite temperatures.
\end{abstract}

\vspace{2pc}

\section{Introduction}

Anomalous Hall effect (AHE) in ferromagnetic materials is one of the representative examples of transport phenomena that reflects quantum nature of electrons in solids~\cite{Nagaosa2006,Sinitsyn2008,Nagaosa2010}. Intensive studies over the last half-century have revealed that the mechanism of AHE reflects rich physics such as the Berry phase of electronic bands~\cite{Karplus1956} and scattering of electrons by impurities; the former is called intrinsic mechanism while the later is extrinsic mechanisms. The extrinsic AHE is a consequence of asymmetric scattering induced by impurities, and many different mechanisms are known such as, non-magnetic impurities~\cite{Smit1955,Smit1958,Berger1970}, localized magnetic moment~\cite{Kondo1962}, and the asymmetric scattering in the Anderson impurity models~\cite{Fert1987,Yamada1993,Kontani1994}. On the other hand, extrinsic AHE by electrons with Rashba spin-orbit interaction (SOI) is also studied, showing that the bulk SOI also contributes to the skew scattering mechanism~\cite{Liu2006,Nunner2007,Kovalev2009}, and it is also proposed that the Berry curvature of electronic bands leads to extrinsic AHE regardless of its microscopic origin~\cite{Adams1959,Ishizuka2017}. In addition, related physics is also studied in relation to extrinsic spin Hall effect~\cite{Engel2005,Guo2008,Gradhand2010,Ferreira2014,Kodderitzsch2015,Yang2016}. While a variety of different mechanisms are proposed, most of them are related to a scattering by single impurity, and the effect of SOI in the scattering process is a crucial ingredient for the anomalous Hall effect, except for that by the Berry curvature.

On the other hand, later studies found that the quantum interference effect induced by multiple scatterers (often by localized magnetic moments) lead to AHE. Unlike the single impurity scattering mechanisms, these mechanisms do not involve SOI in the scattering process. One of such mechanism was theoretically proposed in the strong Kondo coupling limit of the Kondo lattice model. It was pointed out that, in this limit, an effective Peierls phase similar to that by the magnetic field is generated by a twist of the localized moments in the realspace~\cite{Ye1999,Ohgushi2000}, which may appear even in antiferromagnets~\cite{Shindou2001,Martin2008,Chen2014}. This phenomena is similar to the intrinsic AHE in a sence that the modification of band structure by magnetic order induces Berry curvature~\cite{Ye1999,Ohgushi2000}. Such contribution to AHE is studied in the context of non-coplanar magnetic textures, such as in pyrochlore~\cite{Taguchi2001,Machida2010} and kagome~\cite{Nakatsuji2015} magnets, and in relation to magnetic skyrmions in manganites~\cite{Matl1998,Jakob1998,Ye1999,Chun2000} and chiral magnets~\cite{Neubauer2009,Yu2010,Kanazawa2011}.

In the weak coupling limit of the Kondo lattice models, on the other hand, three-spin scattering process gives rise to AHE proportional to scalar spin chirality $\bm S_1\cdot\bm S_2\times\bm S_3$~\cite{Tatara2002} (Scalar spin chirality in Table~\ref{tab:ahe}). It was later discussed that this mechanism is a distinct mechanism from the effective Peierls phase and is rather similar to the skew scattering but by multiple spins instead of non-magnetic impurities~\cite{Ishizuka2018}. Experimentally, this multi-spin skew scattering was studied in relation to disordered spin systems such as chiral spin glass~\cite{Kawamura2003}, chiral magnets~\cite{Kanazawa2011}, and for the skyrmions living on the ferromagnet/semiconductor interfaces~\cite{Denisov2016}. In general, these mechanism are expected to appear in magnets with non-coplanar magnetic orders/correlation. Different microscopic mechanisms for the extrinsic AHE is summarized in Table.~\ref{tab:ahe}; they are classified by the nature of the conduction electrons and impurities. The skew scattering and side-jump mechanisms belongs to the left-bottom quadrant as well as the skew scattering by Rashba SOI and impurity scattering in presence of Berry curvature; the mechanisms listed in green require SOI while that in yellow and blue doesn't. The skew scattering by magnetic impurities (both that by single impurity or by scalar spin chirality) belongs to the right-top quadrant. 

\begin{table*}[bht]
  \begin{center}
  \includegraphics[width=0.75\textwidth]{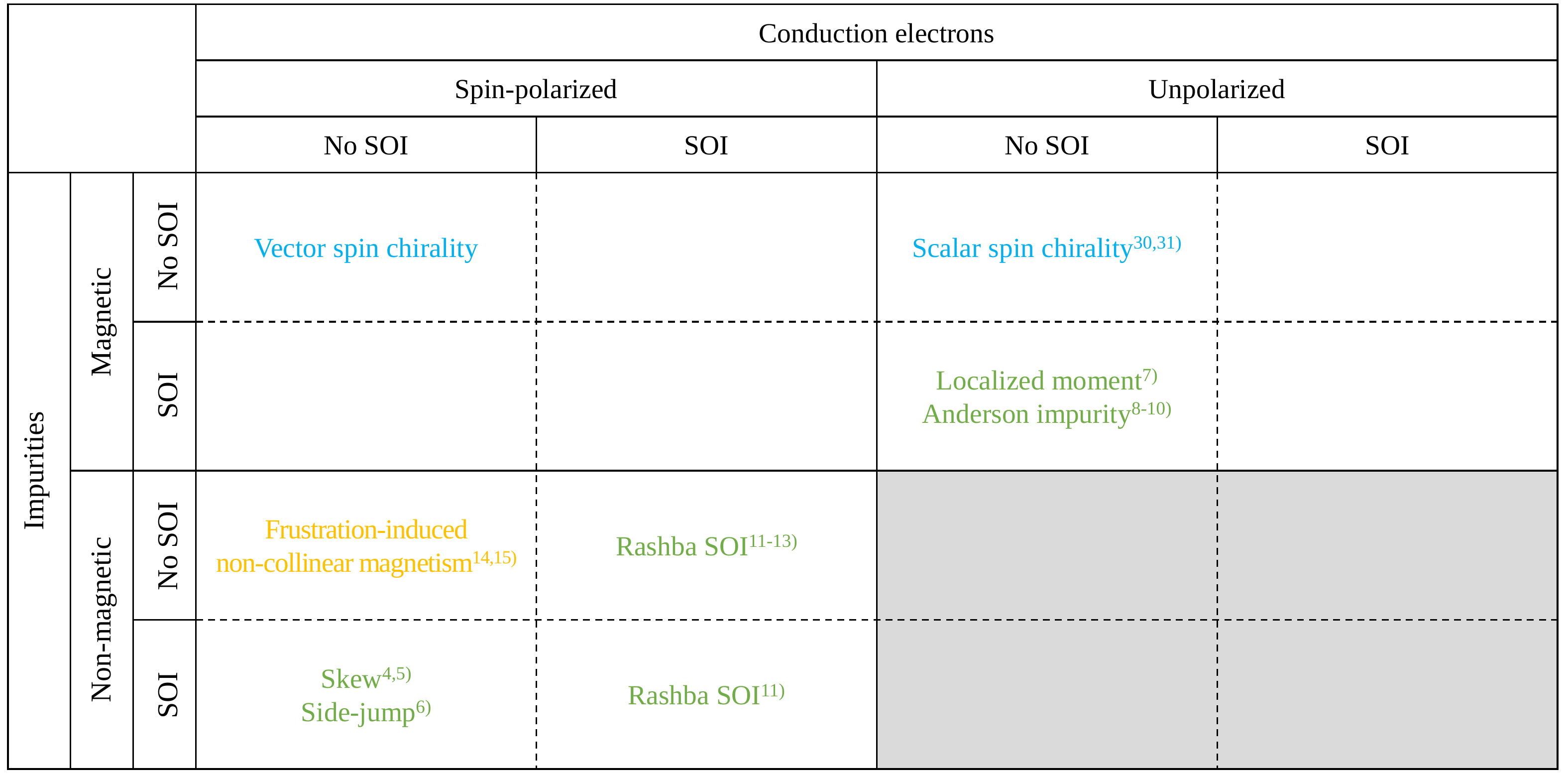}
  \caption{
  Classification of the mechanisms of extrinsic anomalous Hall effect based on the nature of scattering process; whether it requires spin-orbit interaction (SOI) and/or magnetism [Spin-polarization in the electron bands is neccesary or not, and whether the impurities are magnetic or not.]. In this table, we did not consider the origin of the magnetic texture, i.e., whether Dyzaloshinskii-Moriya interaction is necessary for non-coplanar magnetic correlation as it is not related to the scattering mechanism. Each columns are for the different nature of conduction electrons and the rows are for the nature of impurities. The non-magnetic, paramagnetic blocks are shaded in gray as the time-reversal symmetry prohibits the Hall effect. Each color shows different nature of the scattering process: mechanisms that SOI is required (green) and multiple scatterers (blue). Neither SOI nor multiple scatterers are necessary for the orange one. This work is listed as ``vector spin chirality'' in the left-top quadrant.
  }\label{tab:ahe}
  \end{center}
\end{table*}

In the weak coupling limit, the possibility of the AHE by two spin scattering which is proportional to vector spin chirality $\bm S_1\times\bm S_2$ is also studied in several setups. It was initially discussed to vanish in a clean system~\cite{Taguchi2009}, but was later discussed to remain finite in presence of non-magnetic impurities~\cite{Zhang2017}. In this work, we extend the theory in Ref.~\cite{Zhang2017}, and systematically studies the skew scattering mechanism induced by the scattering process involving both spins and non-magnetic impurities. Using second Born approximation, we discuss that the skew scattering appears from the correlation between the non-magnetic impurity and vector spin chirality of spins surrounding the impurity. This mechanism is in contrast to the mechanism studied in Ref.~\cite{Taguchi2009}; Ref.~\cite{Taguchi2009} studies AHE related to the uniform vector spin chirality, while the mechanism studied here is related to the vector spin chirality defined clockwise surrounding the non-magnetic impurity. We also discuss that the vector spin chirality naturally appears in ferromagnets at finite temperature when a charged non-magnetic impurity exists due to the Dzyaloshinskii-Moriya (DM) interaction induced by the impurity; this also gives rise to the correlation between the impurity and the spins. Unlike the scalar spin chirality mechanism~\cite{Tatara2002,Denisov2016,Ishizuka2018} which requires an intrinsic mechanism for a finite scalar spin chirality such as DM interaction or geometrical frustration, this mechanism is possible in ferromagnetic metals on centrosymmetric metals. These results suggest that the vector-chirality-induced AHE may exists universally in the ferromagnetic metals. For instance, this mechanism is potentially relevant to the AHE observed in SrCoO$_3$ (Ref.~\cite{Zhang2017}).

This paper is organized as follows. In Sec.~\ref{sec:model}, we introduce the Kondo lattice model with non-magnetic impurities which we use to study the scattering process. Using the Kondo lattice model, the scattering process that involves two spins and an impurity is studied in Sec.~\ref{sec:skew}. We show that it gives rise to several different antisymmetric scattering terms; one of them is analogous to the skew scattering which is proportional to the vector spin chirality defined anticlockwise around the impurity. In addition, an explicit formula for transverse conductivity is given. In Sec.~\ref{sec:spin}, we discuss that the vector spin chirality generally appears in ferromagnets whenever there is a charged impurity; this is related to the DM interaction induced by the impurity. By a perturbation calculation, we calculate the temperature dependence of the vector spin chirality in the low temperature much below the magnetic transition temperature; we show that a finite vector spin chirality appears at least in the finite temperature, even when the DM interaction is perturbatively small. Sec.~\ref{sec:summary} is devoted to summary and discussions.

\section{Model}\label{sec:model}

To study how the scattering by impurities and magnetic moments affects transport phenomena, we consider a classical spin Kondo lattice model in the ferromagnetic phase with an impurity,
\begin{eqnarray}
  H=H_0+H_K+H_i+H_S
\end{eqnarray}
where
\begin{eqnarray}
  H_0=\sum_{\bm k,\sigma} \varepsilon_{\bm k\sigma}c_{\bm k\sigma}^\dagger c_{\bm k\sigma},
\end{eqnarray}
is the free fermion part of Hamiltonian,
\begin{eqnarray}
  H_K=J_K\sum_{j} \bm S_j \cdot \left\{c_{\alpha}^\dagger(\bm R_j) \bm\sigma_{\alpha\beta} c_{\beta}(\bm R_j)\right\}.\label{eq:Hamil}
\end{eqnarray}
is the Kondo coupling to the localized moments, and
\begin{eqnarray}
H_i=\sum_l V_lc_{\bm R_l\sigma}^\dagger c_{\bm R_l\sigma},
\end{eqnarray}
is the coupling to impurities. $H_S=H_S(\{\bm S_i\})$ is the Hamiltonian for the localized moments, which we will introduce in Sec.~\ref{sec:spin}. Here, $c_{\bm k\sigma}$ ($c_{\bm k\sigma}^\dagger$) is annihilator (creator) for fermions with momentum $\bm k$ and spin $\sigma=\uparrow,\downarrow$ and $\bm S_j$ is the localized moment at $\bm R_j$. The coefficient $\varepsilon_{\bm k\sigma}=\varepsilon_{\sigma}(k)=k^2/(2m)-\Delta\sigma$ is the eigenenergy of the free electrons with mass $m$, momentum $\bm k$ and spin $\sigma$ ($k\equiv|\bm k|$); $V$ is the strength of the impurity potential, and $J_K$ is the Kondo coupling between the localized moments and itinerant electrons. As we are interested in the ferromagnetic phase, we introduced the Zeeman shift $\Delta$ to introduce the population difference of up spin and down spin electrons.

\section{Skew scattering by vector spin chirality}\label{sec:skew}

In this section, we study how the scattering by both impurities and spins contribute to anomalous Hall effect in ferromagnets. In Sec.~\ref{sec:born}, we study the scattering process that involves both impurities and localized spins. We show that the terms that appear from the second-order Born approximation generally induces asymmetric scattering of spins. In Sec.~\ref{sec:skew_imp}, we discuss basic aspects of the scattering term by considering a simple model. We show that, in this limit, the scattering term shows similar feature to the skew scattering by impurities with spin-orbit interaction. The contribution of this term to the anomalous Hall effect is studied in Sec.~\ref{sec:sigmaxy}.

\subsection{Second Born approximation}\label{sec:born}

To study how the spins and impurities affect the transport phenomena in ferromagnetic metals, we first study the scattering by the localized moments and impurities using the Born approximation. Considering $H_i$ and $H_K$ in Eq.~\eqref{eq:Hamil} as the perturbation terms, we consider the terms up to second order in the Born approximation. Within the approximation, we find the first order term as
\begin{eqnarray}
  F^{(1)}(\bm k'\beta,\bm k\alpha) = \frac{J_K}{(2\pi)^3}&\sum_i \bm S_i\cdot\bm \sigma_{\beta\alpha} e^{-i\bm R_i\cdot(\bm k'-\bm k)}+\frac1{(2\pi)^3}\sum_l V_l e^{-i\bm R_l\cdot(\bm k'-\bm k)},
\end{eqnarray}
and the second-order terms as,
\begin{eqnarray}
  F^{(2)}(\bm k'\beta,\bm k\alpha)=\nonumber\\
  -\frac{J_K^2m}{(2\pi)^4}\sum_{i,j}\frac{e^{i\bm k\cdot\bm R_j-i\bm k'\cdot\bm R_i}}{2\delta_{ij}}\left[S_i^z \bm S_j\cdot\bm\sigma_{\beta\alpha}-\sigma_{\beta\alpha}^z\bm S_i\cdot\bm S_j\right] \left(e^{ik_\uparrow\delta_{ij}}-e^{ik_\downarrow\delta_{ij}}\right)\nonumber\\
  -i\frac{J_K^2m}{(2\pi)^4}\sum_{i,j}\frac{e^{i\bm k\cdot\bm R_j-i\bm k'\cdot\bm R_i}}{2\delta_{ij}}\bm\sigma_{\alpha\beta}\cdot\bm S_i\times\bm S_j\left(e^{ik_\uparrow\delta_{ij}}+e^{ik_\downarrow\delta_{ij}}\right)\nonumber\\
  -\frac{J_Km}{(2\pi)^4}\sum_{i,l}\frac{V_l\bm S_i\cdot\bm\sigma_{\beta\alpha}}{\delta_{ij}} \left(e^{i\bm k\cdot\bm R_l-i\bm k'\cdot\bm R_i+ik\delta_{il}}-e^{i\bm k\cdot\bm R_i-i\bm k'\cdot\bm R_l+ik'\delta_{il}}\right).
\end{eqnarray}
\begin{figure}
\includegraphics[width=\linewidth]{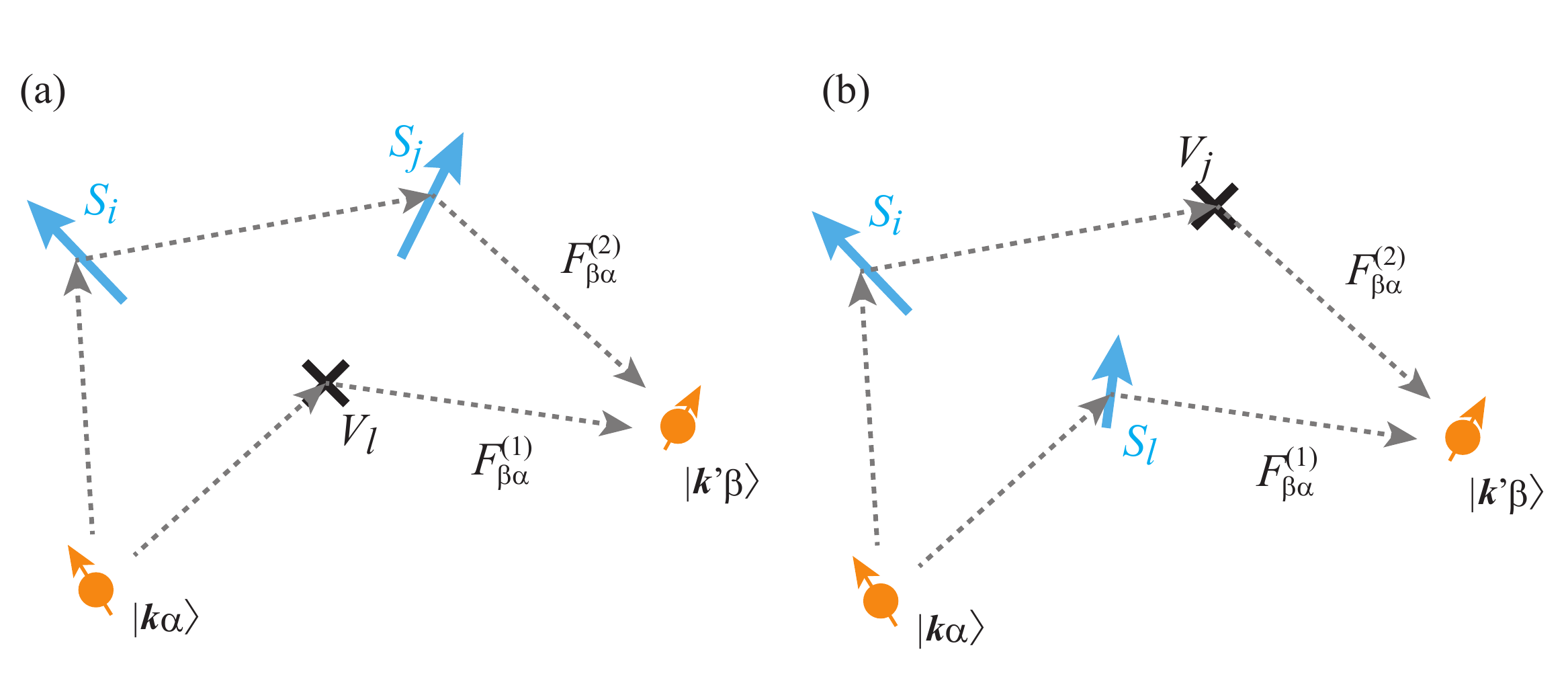}
\caption{
Schematic picture of the scattering processes we consider in this paper. The asymmetric scattering in the third order scattering process of ${\cal O}(J^2V)$ appears from the two processes shown in (a) and (b). The blue arrows show localized moments and the crosses are the non-magnetic impurities.
}
\label{fig:model}
\end{figure}
Here, $\bm \delta_{ij}\equiv \bm R_i-\bm R_j$ and $\delta_{ij}\equiv |\bm R_i-\bm R_j|$ is the distance between the two scatterers. In $F^{(1)}(\bm k'\beta,\bm k\alpha)$, the first term is the contribution from the spin scattering and the second term is by the impurities. Similarly, in $F^{(2)}(\bm k'\beta,\bm k\alpha)$, the first two terms come from the scattering by two spins [See the $F^{(2)}(\bm k'\beta,\bm k\alpha)$ path in Fig.~\ref{fig:model}(a).], and the last term is the scattering by a spin and an impurity [$F^{(2)}(\bm k'\beta,\bm k\alpha)$ path in Fig.~\ref{fig:model}(b).].

In the Born approximation, the probability of scattering electrons with momentum $\bm k$ and spin $\alpha$ ($\left|\bm k\alpha\right>$) to that with $\bm k'$ and $\beta$ ($\left|\bm k'\beta\right>$) reads
\begin{eqnarray}
  W_{\bm k\alpha\to\bm k'\beta}\sim&2\pi|F^{(1)}(\bm k',\bm k)|^2\delta(\varepsilon_{\bm k\alpha}-\varepsilon_{\bm k'\beta})\nonumber\\
  &\quad+2\pi\left(F^{(1)}(\bm k',\bm k)F^{(2)}{}^\ast(\bm k',\bm k) + {\rm h.c}\right)\delta(\varepsilon_{\bm k\alpha}-\varepsilon_{\bm k'\beta})
\end{eqnarray}
Here, the first term is the contribution from the first Born approximation, which only induces the symmetric scattering. On the other hand, the second term is the additional contribution that appears in the second Born approximation. The second Born contributions include terms proportional to $J_K^3$, $VJ_K^2$, $V^2J_K$, and $V^3$. The contribution from the terms that are proportional to $J_K^3$ in the paramagnetic case ($\Delta=0$) is already studied in preceding theoretical works~\cite{Tatara2002,Ishizuka2018}; the anomalous Hall effect appears when the thermal average of scalar spin chirality $\bm S_i\cdot\bm S_j\times\bm S_l$ is finite, such as in chiral spin grass~\cite{Tatara2002,Kawamura2003} and in chiral magnets~\cite{Ishizuka2018}. 

In this work, we focus on the contribution from the impurities, focusing on the leading order term, i.e., ${\cal O}(VJ^2)$ terms. As we are interested in the Hall effect, we further split the scattering term into symmetric [$W_{\bm k\alpha\to\bm k'\beta}^+=(W_{\bm k\alpha\to\bm k'\beta}+W_{\bm k'\beta\to\bm k\alpha})/2$] and asymmetric parts [$W_{\bm k\alpha\to\bm k'\beta}^-=(W_{\bm k\alpha\to\bm k'\beta}-W_{\bm k'\beta\to\bm k\alpha})/2$]. The asymmetric part of the scattering appears from the three different contributions
\begin{eqnarray}
  W_{\bm k\alpha\to\bm k'\beta}^-= W^{(a)}_{\bm k\alpha\to\bm k'\beta} + W^{(a')}_{\bm k\alpha\to\bm k'\beta} + W^{(b)}_{\bm k\alpha\to\bm k'\beta}.
\end{eqnarray}
These contributions are schematically shown in Fig.~\ref{fig:model}. The first two terms $W^{(a)}_{\bm k\alpha\to\bm k'\beta}$ and $W^{(a')}_{\bm k\alpha\to\bm k'\beta}$ come from the interference of second order process involving two spins and first-order process for impurities as shown in Fig.~\ref{fig:model}(a). These terms read
\begin{eqnarray}
  &W^{(a)}_{\bm k\alpha\to\bm k'\beta}=\frac{J_K^2m}{(2\pi)^7}\sum_{i,j,l,\gamma} \frac{V_l\sin(k_\gamma\delta_{ij})}{2\delta_{ij}} \nonumber\\
  &\quad\times\left[\bm \sigma_{\alpha\beta}\cdot\bm S_i\times\bm S_j\left\{\cos(\bm k\cdot\bm\delta_{jl}-\bm k'\cdot\bm\delta_{il})-\cos(\bm k\cdot\bm\delta_{il}-\bm k'\cdot\bm\delta_{jl})\right\}\right.\nonumber\\
  &\quad-\left.{\rm sgn}(\gamma)\sigma_{\alpha\beta}^z(S_i^x S_j^x+S_i^yS_j^y)\left\{\sin(\bm k\cdot\bm\delta_{jl}-\bm k'\cdot\bm\delta_{il})+\sin(\bm k\cdot\bm\delta_{il}-\bm k'\cdot\bm\delta_{jl})\right\}\right],\nonumber\\
\label{eq:W1}\\
  &W^{(a')}_{\bm k\alpha\to\bm k'\beta}=-\frac{J^2m}{(2\pi)^7}\sum_{i,j,l,\alpha,\beta} \frac{V_l}{2\delta_{ij}}\left\{S_i^z(S_j^x\sigma_{\alpha\beta}^x+S_j^y\sigma_{\alpha\beta}^y)\right\}\nonumber\\
&\qquad\times\left[\cos(\bm k\cdot\bm\delta_{jl}-\bm k'\cdot\bm\delta_{il}+k\delta_{ij})+\cos(\bm k\cdot\bm\delta_{jl}-\bm k'\cdot\bm\delta_{il}+k'\delta_{ij})\right.\nonumber\\
&\qquad\qquad\left.-\cos(\bm k'\cdot\bm\delta_{jl}-\bm k\cdot\bm\delta_{il}+k\delta_{ij})-\cos(\bm k'\cdot\bm\delta_{jl}-\bm k\cdot\bm\delta_{il}+k'\delta_{ij})\right].
\end{eqnarray}
Here, ${\rm sgn}(\gamma)$ is a binary function that is ${\rm sgn}(\uparrow)=1$ and ${\rm sgn}(\downarrow)=-1$ and $k_\gamma\in\mathbb R$ ($\gamma=\uparrow,\downarrow$) is the magnitude of the wave number such that $\varepsilon_\uparrow({k_\uparrow})=\varepsilon_\downarrow({k_\downarrow})$. The second term $W^{(b)}_{\bm k\alpha\to\bm k'\beta}$ comes from second order process involving the scattering by a spin and an impurity, and the first-order scattering term by the spins [For example, the process shown in Fig.~\ref{fig:model}(b)]. This term reads
\begin{eqnarray}
  W^{(b)}_{\bm k\alpha\to\bm k'\beta}&=&-i\frac{J_K^2m}{(2\pi)^7}\sum_{i,j,l}\frac{V_l}{\delta_{il}}(\bm S_j\cdot\bm\sigma_{\alpha\beta})(\bm S_i\cdot\bm\sigma_{\beta\alpha})\times\nonumber\\
 &&\left\{e^{ik\delta_{il}}\sin(\bm k\cdot\bm \delta_{lj}-\bm k'\cdot\bm\delta_{ij})+e^{ik'\delta_{il}}\sin(\bm k\cdot\bm \delta_{ij}-\bm k'\cdot\bm\delta_{lj})\right\}\nonumber\\
 &&+{\rm h.c.}.\label{eq:W2}
\end{eqnarray}
In general, both these terms contribute to asymmetric scattering and to the anomalous Hall effect.

In the next section, however, we find that only the terms with $\bm S_i\times\bm S_j$ in $W^{(a)}_{\bm k\alpha\to\bm k'\beta}$ gives rise to skew scattering while the other terms vanish in a most basic setup. In particular, assuming $k\delta_{il}, k\delta_{jl}\ll1$ and the net magnetization along the $z$ axis, we find
\begin{eqnarray}
  W^{(a)}_{\bm k\alpha\to\bm k'\beta}&=&\frac{J_K^2m}{(2\pi)^7}\sum_{i,j,l,\gamma} \frac{V_l\sin(k_\gamma\delta_{ij})}{2\delta_{ij}} \sigma^z_{\alpha\beta}\left(\hat{\bm z}\cdot\bm S_i\times\bm S_j\right)\nonumber\\
  &&\times\left\{\cos(\bm k\cdot\bm\delta_{jl}-\bm k'\cdot\bm\delta_{il})-\cos(\bm k\cdot\bm\delta_{il}-\bm k'\cdot\bm\delta_{jl})\right\}\\
&\sim&\frac{J_K^2m}{(2\pi)^7}\sum_{i,j,l,\gamma} \frac{V_l k_\gamma}{2} \sigma^z_{\alpha\beta}\left(\hat{\bm z}\cdot\bm S_i\times\bm S_j\right)\left[(\bm k\times\bm k')\cdot(\bm\delta_{ji}\times\bm\delta_{ij;l})\right]\nonumber\\
  &&+\frac{J_K^2m}{(2\pi)^7}\sum_{i,j,l,\gamma} \frac{V_l k_\gamma}{2} \sigma^z_{\alpha\beta}\left(\hat{\bm z}\cdot\bm S_i\times\bm S_j\right)\nonumber\\
&&\qquad\times\left[(\bm k\cdot\bm\delta_{ij;l})(\bm k\cdot\bm\delta_{ij})+(\bm k'\cdot\bm\delta_{ij;l})(\bm k'\cdot\bm\delta_{ji})\right].\label{eq:W1smallk}
\end{eqnarray}
Here, we assumed that $S_i^x, S_i^y\ll S_i^z\sim 1$ and only considered the leading order terms in $S_i^x$ and $S_i^y$. The first term in the second line gives a scattering term proportional to $\bm k\times\bm k'$, which is a typical wave-number dependence for the skew scattering~\cite{LerouxHugon1972,Bruno2001}. Here, $\bm\delta_{ij;l}\equiv(\bm\delta_{il}+\bm\delta_{jl})/2$ is a vector from the non-magnetic impurity to the bonds connecting $\bm S_i$ and $\bm S_j$ [Fig.~\ref{fig:polarization}(a)]. The result indicates that the correlation between the spins and non-magnetic impurity is crucial for the skew scattering. Another important fact is that the skew scattering is proportional to $\langle V_l\left(\hat{\bm z}\cdot\bm S_i\times\bm S_j\right)(\bm\delta_{ji}\times\bm\delta_{ij;l})\rangle$ where the brackets are for thermal and impurity averages; the skew scattering is proportional to the skew scattering defined anticlockwise around the impurity, and not to the uniform vector spin chirality which is finite in helical magnetic states.

\subsection{Scattering by an impurity and surrounding moments}\label{sec:skew_imp}

\begin{figure}
\includegraphics[width=\linewidth]{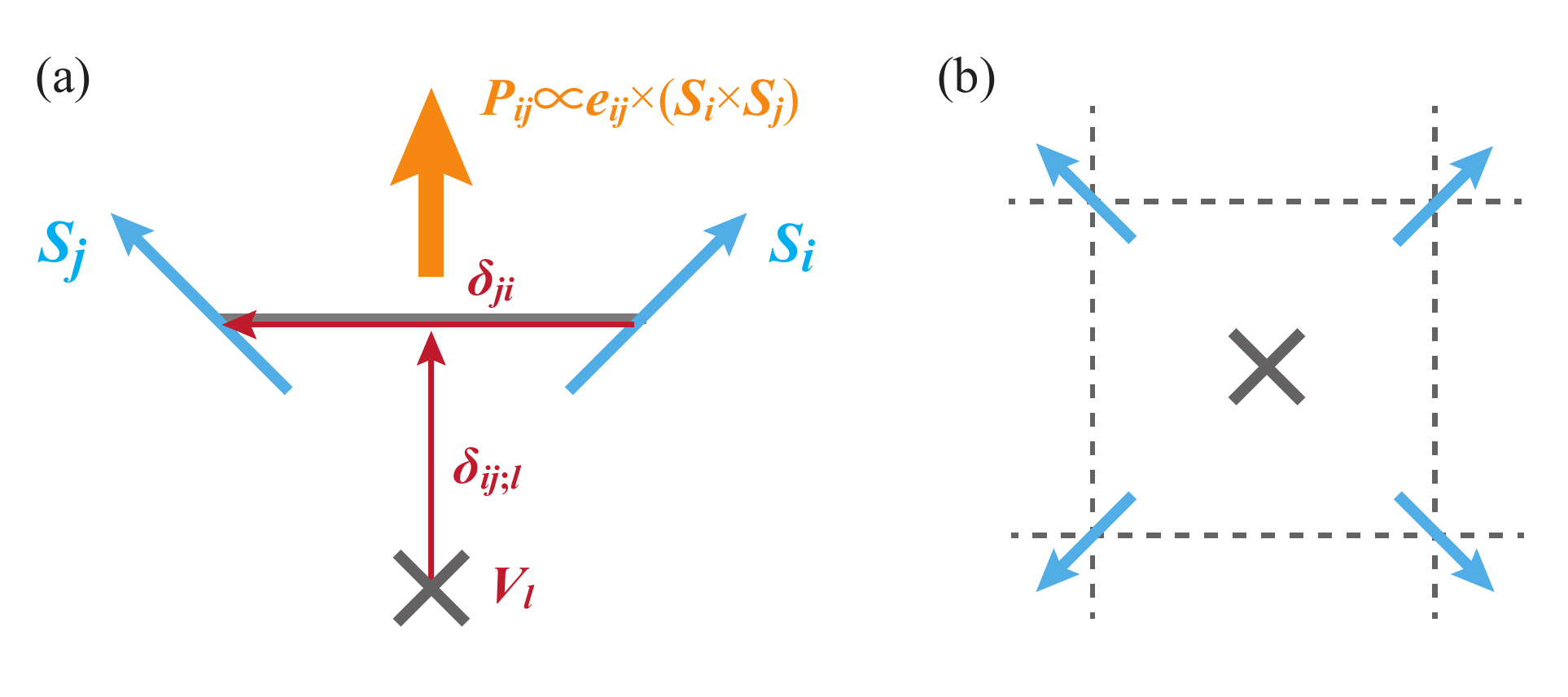}
\caption{
  (a) Schematic figure of the electric polarization induced by the canting of two spins $\bm P_{ij}$, and (b) the model we consider as an example.
}
\label{fig:polarization}
\end{figure}
 
To see how the above mentioned mechanism contribute to the transport phenomena, we here consider a basic model with an impurity and a plaquette of four moments surrounding the impurity. A schematic picture of the model considered is shown in Fig.~\ref{fig:polarization}(b). For simplicity, we assume the impurity is at $\bm R_0=\bm 0$ and the four surrounding moments are at
\begin{eqnarray*}
  {\bm R}_1=(\frac{a}2,-\frac{a}2,0),\quad {\bm R}_2=(\frac{a}2,\frac{a}2,0),\nonumber\\
  {\bm R}_3=(-\frac{a}2,\frac{a}2,0),\quad {\bm R}_4=(-\frac{a}2,-\frac{a}2,0).
\end{eqnarray*}
We further assume that the magnetic moment is nearly polarized along $z$ axis with small fluctuation for $S_i^x$ and $S_i^y$, and that $\langle(\bm S_i\times\bm S_{i+1})_z\rangle=\bar c$ for all $i$ and $\langle(\bm S_i\times\bm S_{i+2})_z\rangle=0$; these assumptions give $\langle V_l\left(\hat{\bm z}\cdot\bm S_i\times\bm S_j\right)(\bm\delta_{ji}\times\bm\delta_{ij;l})\rangle=-V_l\bar ca^2/2$. As we will discuss in the next section, this is a natural assumption for the current model. Using these assumptions and expanding the Eqs.~\ref{eq:W1} and \ref{eq:W2} by $\bm k$ in the limit of $ka\ll1$, we find
\begin{eqnarray}
  W^{(a)}_{\bm k\alpha\to\bm k'\beta}&=-\frac{2J_K^2V_0ma^2}{(2\pi)^7}\bar c(k_{\uparrow}+k_{\downarrow})(\bm k\times\bm k')_z\sigma^z_{\alpha\beta},\label{eq:W1simple}\\
  W^{(a')}_{\bm k\alpha\to\bm k'\beta}&=0,\\
  W^{(b)}_{\bm k\alpha\to\bm k'\beta}&=0.
\end{eqnarray}
Here, $W^{(a')}_{\bm k\alpha\to\bm k'\beta}$ vanish as they only contribute to higher-order terms when $\langle S_i^x\rangle=\langle S_i^y\rangle=0$. The results indicate that the scattering process we considered give rise to an asymmetric scattering term analogous to the skew scattering, i.e., its wave-number dependence proportional to $\bm k\times\bm k'$. This indicates that only the $\bm S_i\times\bm S_j$ terms in Eq.~\ref{eq:W1smallk} contributes to the skew scattering.

\subsection{Anomalous Hall effect}\label{sec:sigmaxy}

We next evaluate the conductivity by using a Boltzmann theory considering the asymmetric scattering term in Eq.~\eqref{eq:W1simple}. We here assume that the above considered impurity-spin cluster is randomly distributed throughout the system with density of $n^{(i)}$. Assuming the system is spatially uniform, the Boltzmann equation reads
\begin{eqnarray}
  q\bm v_{\bm k}\cdot\bm E f'_0(\varepsilon_{\bm k\sigma}) = -\frac{g_{\bm k\sigma}}{\tau}+\sum_\beta\int d\phi' d\theta' \sin\theta' \frac{\rho(k)}{4\pi} \hat V_\sigma\cdot\frac{\bm k\times\bm k'}{k^2} g_{\bm k'\sigma},\label{eq:boltzmann_main}
\end{eqnarray}
where
\begin{eqnarray}
  {\hat V}_\sigma(k) = -2\,{\rm sgn}(\sigma)n^{(i)}V_0J_K^2ma^2\bar ck_\sigma^2(k_{\uparrow}+k_{\downarrow}).\label{eq:Vsk}
\end{eqnarray}
Here, $\rho(k)=mk/(2\pi^2)$ is the density of state for the wavenumber $k$. In the right-hand side of the equation, the first term is the symmetric scattering term which we approximated by relaxation time approximation, and the second term is the asymmetric scattering term in Eq.~\eqref{eq:W1simple}.

The Boltzmann equation in Eq.~\eqref{eq:boltzmann_main} can be solved without further approximation~\cite{Ishizuka2017,Ishizuka2018}. The result reads
\begin{eqnarray}
  \bm j &= \bm j_\uparrow + \bm j_\downarrow,\\
  \bm j_\sigma &= \frac{nq^2\tau}m\left(\bm E - 2\pi \tau {\hat V}_\sigma(k_{F\sigma})\times \bm E\right).\label{eq:js}
\end{eqnarray}
Here, $k_{F\sigma}$ is the Fermi wavenumber for the electrons with spin $\sigma$. Hence, we obtain
\begin{eqnarray}
\sigma_{xx}&=\frac{q^2\tau}m(n_\uparrow+n_\downarrow),\nonumber\\
\sigma_{xy}&=\frac{q^2\tau^2}{2m} [\rho(k_{F\uparrow})V_\uparrow(k_{F\uparrow}) n_\uparrow +\rho(k_{F\downarrow})V_\downarrow(k_{F\downarrow}) n_\downarrow],
\end{eqnarray}
for the longitudinal and transverse conductivities, respectively.

The Hall conductivity is linearly proportional to $V_\sigma(k)\propto\langle V_l\left(\hat{\bm z}\cdot\bm S_i\times\bm S_j\right)(\bm\delta_{ji}\times\bm\delta_{ij;l})\rangle$, the vector spin chirality of the localized moments defined anticlockwise around the impurity. The result is linearly proportional to $V_0$, the impurity potential, therefore no AHE occurs if there are no correlation between the non-magnetic impurity and the spin-spin correlation, i.e., if $\langle V_l\left(\hat{\bm z}\cdot\bm S_i\times\bm S_j\right)(\bm\delta_{ji}\times\bm\delta_{ij;l})\rangle=\langle V_l\rangle\langle\left(\hat{\bm z}\cdot\bm S_i\times\bm S_j\right)(\bm\delta_{ji}\times\bm\delta_{ij;l})\rangle$, as $\langle V_l\rangle=0$. We also note that the contribution from the up spin and down spin electrons have opposite sign as shown in Eq.~\eqref{eq:Vsk}. Therefore, the Hall effect is absent when there is no Zeeman splitting $\Delta=0$. This implies that the leading order in the anomalous Hall conductivity is proportional to the magnetization.

\section{Temperature dependence of Hall conductivity}\label{sec:spin}

We next study how the AHE studied in the above sections behave at finite temperature. In Sec.~\ref{sec:spin:dm}, we show that the colinear ferromagnetic state is perturbatively stable against the DM interaction induced by the impurity. Therefore, no AHE appears in the $T\to0$ limit. Nevertheless, we show that the expectation for vector spin chirality become finite in the finite temperature due to the spin fluctuation. In Sec.~\ref{sec:spin:chirality}, we introduce the formalism we use which is based on the spin-wave theory. Using the formalism, we discuss the temperature dependence of the vector spin chirality.

\subsection{Impurity-induced Dzyaloshinskii-Moriya interaction}\label{sec:spin:dm}

In this section, we discuss how the non-magnetic impurities affect the magnetic ground state. As the spin model, we consider a ferromagnetic Heisenberg model on the cubic lattice with an impurity. The Hamiltonian reads
\begin{eqnarray}
  H_S = H_S^{(0)} + H_S^{(1)}
  \label{eq:Hs}
\end{eqnarray}
where
\begin{eqnarray}
  H_S^{(0)}=-J\sum_{\langle \bm R,\bm R'\rangle}\bm S_{\bm R}\cdot\bm S_{\bm R'}-h\sum_{\bm R}S^z_{\bm R},
  \label{eq:Hs:heis}
\end{eqnarray}
is the Heisenberg Hamiltonian. The first term is Heisenberg exchange interaction between the localized moments where $\bm S_{\bm R}$ is the Heisenberg spin on $\bm R=(n_x+1/2,n_y+1/2,n_z)$ site [$n_\alpha\in\mathbb Z$ ($\alpha=x,y,z$)], and $J>0$ is the exchange coupling between the nearest-neighbor spins; the sum is over nearest-neighbor sites. The second term is the coupling to the external magnetic field which we assume to be along $z$ axis.

In the presence of an impurity, the spins surrounding the impurity are affected by the electric potential by the impurity. The spins couples to the electric potential via electric dipole moments induced by spin canting~\cite{Katsura2005}, which is proportional to $\bm P_{\bm R,\bm R'}\propto \bm\delta_{\bm R',\bm R}\times(\bm S_{\bm R}\times\bm S_{\bm R'}),$ where $\bm\delta_{\bm R',\bm R}=\bm R'-\bm R$ is the vector connecting $\bm S_{\bm R}$ and $\bm S_{\bm R'}$. Therefore, the impurity term of Hamiltonian reads
\begin{eqnarray}
  H_S^{(imp)}=-\frac12\sum_{\bm R',\bm R} (\hat D_{\bm R',\bm R}\times\bm\delta_{\bm R',\bm R})\cdot\left(\bm S_{\bm R}\times\bm S_{\bm R'}\right).
\end{eqnarray}
Here, $\hat D_{\bm R',\bm R}$ is the coupling vector proportional to the direction of the electric field (or the gradient of electric potential) at the bond center of $\bm S_{\bm R}$ and $\bm S_{\bm R'}$. For simplicity, we here assume the impurity is at $\bm R=\bm 0$ and consider only the coupling to the four spins surrounding the impurity [See Fig.~\ref{fig:polarization}(b)]:
\begin{eqnarray}
  {\bm R}_1=(\frac12,-\frac12,0),\quad {\bm R}_2=(\frac12,\frac12,0),\quad {\bm R}_3=(-\frac12,\frac12,0),\quad {\bm R}_4=(-\frac12,-\frac12,0).\nonumber\\
\end{eqnarray}
Under this approximation, the impurity term simplifies to
\begin{eqnarray}
  H_S^{(1)}=-D\sum_{i=1}^4 [(\bm R_i+\bm R_{i+1})\times\bm\delta_{\bm R_{i+1},\bm R_i}]\cdot\left(\bm S_{\bm R_i}\times\bm S_{\bm R_{i+1}}\right).
\end{eqnarray}
Here, $D$ is the strength of coupling and ${\bm R}_5={\bm R}_1$; we assume the electric potential by the impurity is rotationally symmetric. The above argument shows that the electric potential due to the impurity induces DM interaction between the surrounding spins. However, unlike the DM interaction in chiral magnets, the ferromagnetic ground state is stable against infinitesimally small DM interaction. In \ref{sec:fmstability}, we discuss that the collinear ferromagnetic order remains as the classical ground state when $|D|<J+h/2$, unlike the DM interaction in chiral magnets. Therefore, we expect no spin canting in the ground state.

\subsection{Temperature dependence of vector spin chirality}\label{sec:spin:chirality}

In contrast to the ground state, at finite temperature, the DM interaction discussed in the previous section could give rise to finite spin chirality due to the lifting of degeneracy between the excitations with positive and negative vector spin chirality. This may gives rise to the finite spin chirality and resulting anomalous Hall effect that appears only at finite temperature. To study whether the weak DM interaction contributes to finite chirality at finite temperature, we focus on the low temperature region well below the magnetic transition temperature. Assuming the spin fluctuation to be small, we study the temperature dependence of spin chirality using a spin wave theory.

By using Holsten-Primakov representation
\begin{eqnarray}
  S^z_i=S-a_i^\dagger a_i,\quad S^-_i=\sqrt{2S}a^\dagger_i\left(1-\frac{a_i^\dagger a_i}{2S}\right)^\frac12,\quad S^+_i=\sqrt{2S}\left(1-\frac{a_i^\dagger a_i}{2S}\right)^\frac12a_i.\nonumber\\
\end{eqnarray}
and leaving the terms up to the order of ${\cal O}(S)$, the Hamiltonian in Eq.~\eqref{eq:Hs} become
\begin{eqnarray}
  H_S &= \sum_{\bm k} \varepsilon_{\bm k} a_{\bm k}^\dagger a_{\bm k}-i\sqrt2DS\sum_{i=1}^4 a^\dagger({\bm R}_{i+1})a({\bm R}_{i})-a^\dagger({\bm R}_{i})a({\bm R}_{i+1}),\nonumber\\
  &= \sum_{\bm k} \varepsilon_{\bm k} a_{\bm k}^\dagger a_{\bm k}+\sum_{{\bm k},{\bm k'}} V_{{\bm k},{\bm k'}}a_{\bm k}^\dagger a_{\bm k'},
\end{eqnarray}
where $\varepsilon_{\bm k}=2JS[3-\sum_{\alpha=x,y,z} \cos(k_\alpha)]+h$ is the eigenenergy of the spin wave mode and
\begin{eqnarray}
  V_{{\bm k},{\bm k'}}=\frac{\sqrt2DS}N\chi_{\bm k,\bm k'},
\end{eqnarray}
with
\begin{eqnarray}
  \chi_{\bm k,\bm k'}=&-i\left\{e^{i(\bm k-\bm k')\cdot R_2}(e^{ik'_y}-e^{-ik_y})+e^{i(\bm k-\bm k')\cdot R_3}(e^{-ik'_x}-e^{ik_x})\right.\nonumber\\
  &\qquad\left.+e^{i(\bm k-\bm k')\cdot R_4}(e^{-ik'_y}-e^{ik_y})+e^{i(\bm k-\bm k')\cdot R_1}(e^{ik'_x}-e^{-ik_x})\right\}.
\end{eqnarray}

\begin{figure}
\centering
\includegraphics[width=0.6\linewidth]{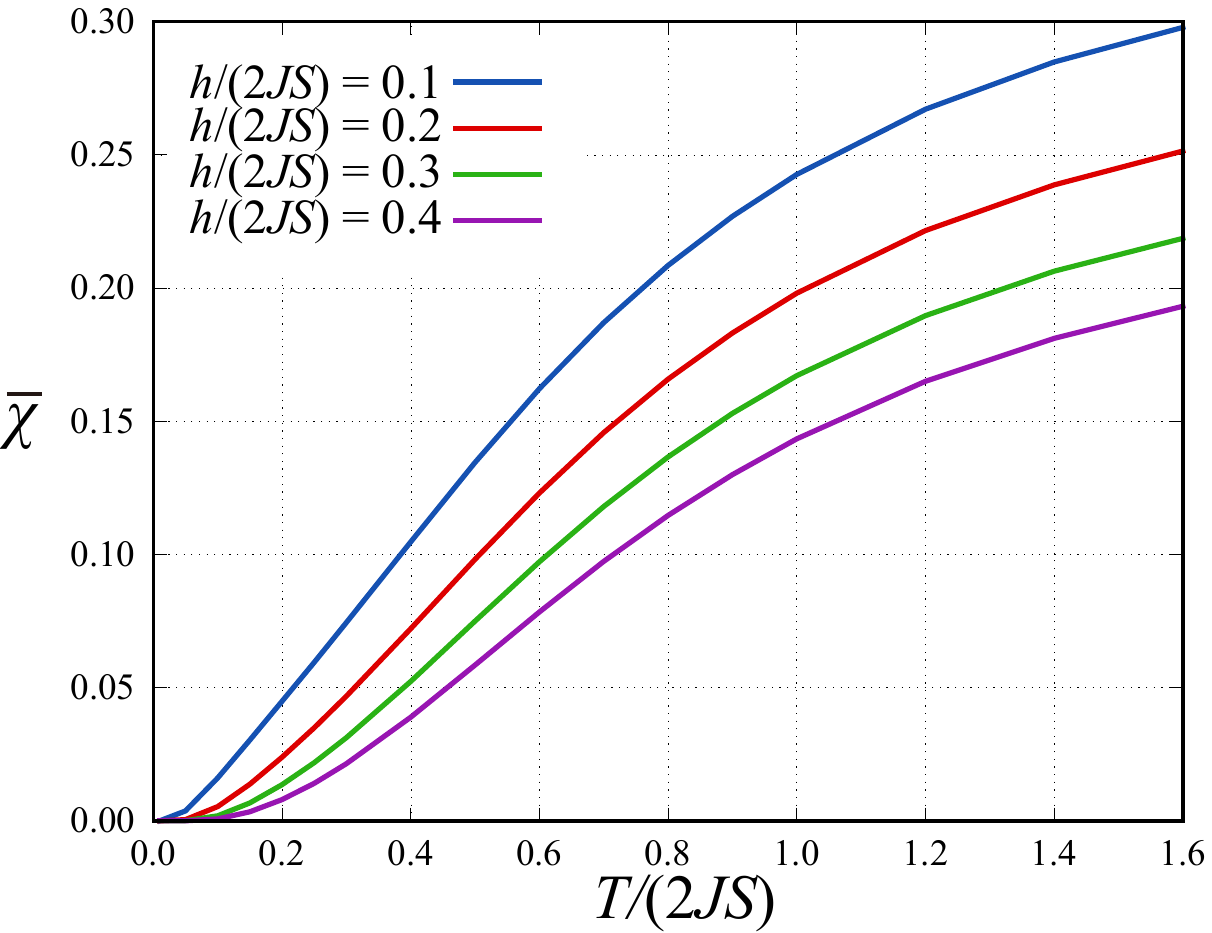}
\caption{
  Temperature dependence of $\bar\chi=\sqrt2J\left<\bm S_1\times\bm S_2\right>/(DS)$ calculated for the different strength of magnetic field $h$. The magnetic field induces a spin gap in the spin-wave modes suppressing the vector spin chirality at the low temperature. The anomalous Hall conductivity is proportional to $\bar\chi$ as shown in Eq.~\eqref{eq:js}, using Eqs.~\eqref{eq:Vsk}, \eqref{eq:chirality}, and \eqref{eq:chibar}.
}
\label{fig:chibar}
\end{figure}

Using the perturbation expansion with respect to $H_S^{(1)}$, the vector spin chirality of spins $\langle \bm S_1\times\bm S_2\rangle$ is given by
\begin{eqnarray}
  \bar c &=& \langle \bm S_1\times\bm S_2\rangle = -\frac{\langle H_{imp}\rangle}{4(D/\sqrt2)}\nonumber\\
  &\sim& \frac1{2\sqrt2D}\sum_{l}e^{i\omega_l\eta}{\rm tr}\left[\hat{V}\hat{G}^0(i\omega_l)\hat{V}\hat{G}^0(i\omega_{l})\right]\nonumber\\
  &=& \frac{DS}{\sqrt2J}\bar\chi,\label{eq:chirality}
\end{eqnarray}
where $\hat{V}$ and $\hat{G}^0$ are the matrix representation of $V_{\bm k,\bm k'}$ and the Matsubara Green's function for bosons $G^0_{\bm k',\bm k}(i\omega_l)\equiv 1/(i\omega_l-H_S^{(0)})$ ($\omega_l=2\pi l/\beta$), respectively. In the last equation,
\begin{eqnarray}
  \bar\chi=-\frac\beta{N^2}\sum_{\varepsilon'_{\bm k}\ne\varepsilon'_{\bm k'}}\frac{|\chi_{\bm k,\bm k'}|^2}{\varepsilon'_{\bm k}-\varepsilon'_{\bm k'}}n_{\bm k}.\label{eq:chibar}
\end{eqnarray}
Here, $n_{\bm k}\equiv1/(e^{\beta\varepsilon_{\bm k}}-1)$ is the Bose distribution function and $\varepsilon'_{\bm k}\equiv\varepsilon_{\bm k}/(2JS)$ is the renormalized energy.

In Fig.~\ref{fig:chibar}, we show the temperature dependence of $\bar\chi$ calculated numerically using Eq.~\eqref{eq:chibar}. The different curves are for the different strength of magnetic field. At a low temperature, typically below $T/(2SJ)\lesssim0.2$, the result show suppression of the chiral spin fluctuation due to the spin gap induced by the external magnetic field. Above it, we see a linear growth of $\bar\chi$, which is expected in the classical approximation up to $T/(2JS)\lesssim 1$. The trends of our results resembles that of the finite temperature component of AHE in SrCoO$_3$~\cite{Zhang2017}, and may also be applicable to other oxides where enhancement of AHE is often observed at a finite temperature.

\section{Discussion and Summary}\label{sec:summary}

To summarize, in this work, we discussed the asymmetric scattering and the anomalous Hall effect induced by the correlation of non-magnetic impurities and spins, which brings about a new mechanism for anomalous Hall effect proportional to the vector spin chirality defined anticlockwise around the impurity. By using a scattering theory approach, we show that the Hall conductivity is proportional to the correlation between the non-magnetic impurity and vector spin chirality, $V_\sigma(k)\propto\langle V_l\left(\hat{\bm z}\cdot\bm S_i\times\bm S_j\right)(\bm\delta_{ji}\times\bm\delta_{ij;l})\rangle$, where $\bm z$ is the direction of the magnetization; this mechanism is interpreted as a skew scattering induced by the quantum phase interference due to the multiple scatterers. We further show that the correlation between the non-magnetic impurity and the vector spin chirality of the spins surrounding the impurity naturally appears due to the multiferroic nature of the spin canting.

An interesting aspect of this mechanism is that the correlation between the non-magnetic impurity and the surrounding localized moments generally appears at a finite temperature, even if it is zero at $T=0$. Therefore, the mechanism contributes to the enhancement of Hall effect at a finite temperature; the temperature dependence of the vector spin chirality is given in Fig.~\ref{fig:chibar}. This mechanism is potentially relevant to the enhancement of anomalous Hall conductivity with increasing temperature found in ferromagnetic oxides.

\section*{Acknowledgements}

The authors thank M. Ok, P. Yu, and D. Zhang for fruitful discussions. This work was supported by JSPS KAKENHI Grant Numbers JP16H06717, JP18H03676, JP18H04222, JP26103006, ImPACT Program of Council for Science, Technology and Innovation (Cabinet office, Government of Japan), and CREST, JST (Grant No. JPMJCR16F1).

\appendix

\section{Stability of the ferromagnetic ground state}\label{sec:fmstability}

To study the stability of the ferromagnetic order in the Hamiltonian in Eq.~\eqref{eq:Hs}, we first divide the spins in two groups: the four spins surrounding the impurity ($A$) and other spins ($B$). Using this grouping, the Hamiltonian is divided into three parts: interactions between the four spins surrounding the impurity ($H_A$), interactions between the other spins ($H_B$), and the interactions between the spins in $A$ and $B$ ($H_{AB}$). The ground state of $H_B$ is obviously the collinear ferromagnetic order and $H_{AB}$ is the ferromagnetic Heisenberg interaction between the spins in $A$ and $B$. Therefore, if the ground state of $H_A$ is the ferromagnetic order, the ground state of the entire system is the collinear ferromagnetic order.

For the Hamiltonian in Eq.~\eqref{eq:Hs}, the term $H_A$ reads
\begin{eqnarray}
H_A=-J\sum_{i=1}^4 \bm S_{\bm R_i}\cdot\bm S_{\bm R_{i+1}}-D\sum_{i=1}^4\left(S_{\bm R_i}^xS_{\bm R_{i+1}}^y-S_{\bm R_i}^yS_{\bm R_{i+1}}^x\right)-h\sum_i S_{\bm R_i}^z.\nonumber\\\label{eq:HA}
\end{eqnarray}
In the second term, we used the fact $(\bm R_i+\bm R_{i+1})\times\bm\delta_{\bm R_{i+1},\bm R_i}=\hat z$, where $\hat z$ is the unit vector along $z$ axis. To study the local stability of the ferromagnetic state, we consider small fluctuation of the spins round the ferromagnetic order along $z$ axis. Namely, we expand $S_{\bm R_i}^z\sim1-[(S_{R_i}^x)^2+(S_{R_i}^y)^2]/2$. The approximated Hamiltonian reads
\begin{eqnarray}
  H_A\sim\sum_{l=0,\gamma=\pm}^3 S_{k_l}^\gamma\left[ J\left\{1-\cos(k_l)\right\}+\frac{h}2 + \gamma D\sin(k_l) \right] \left(S_{k_l}^\gamma\right)^\ast,
\end{eqnarray}
where $k_l=\pi l/2$ and
\begin{eqnarray*}	
  S_{k_l}^\gamma = \frac1{2\sqrt2}\sum_{j=1}^4 \left(S_{\bm R_j}^x - \gamma iS_{\bm R_j}^y\right)e^{-ik_l j}.
\end{eqnarray*}
As the $S_{k_0}^\gamma\ne 0$ solutions are also a collinear ferromagnetic state, the ground state remains the ferromagnetic order for $J+h/2>|D|$.

\end{document}